%% file: Zwier-SiC-CPT.tex
\documentclass[aps,prb,superscriptaddress,preprint]{revtex4}
\usepackage{amsmath}
\usepackage{amssymb}
\usepackage{graphicx}
\usepackage{braket}

\begin{document}



\makeatletter
 \renewcommand\@biblabel[1]{#1.}
\makeatother

\title{All-optical coherent population trapping with\\
defect spin ensembles in silicon carbide}

\author{Olger~V.~Zwier}
\author{Danny~O'Shea}
\author{Alexander~R.~Onur}
\author{Caspar~H.~van~der~Wal}
\affiliation{Zernike Institute for Advanced Materials, University of Groningen, NL-9747AG  Groningen, The Netherlands}

\date{Version of \today}

\begin{abstract}
Divacancy defects in silicon carbide have long-lived electronic spin states and sharp optical transitions, with properties that are similar to the nitrogen-vacancy defect in diamond. We report experiments on 4H-SiC that investigate all-optical addressing of spin states with the zero-phonon-line transitions. Our magneto-spectroscopy results identify the spin $S=1$ structure of the ground and excited state, and a role for decay via intersystem crossing. We use these results for demonstrating coherent population trapping of spin states with divacancy ensembles that have particular orientations in the SiC crystal.
\end{abstract}

\maketitle


Strong interaction between a long-lived spin state and an optical field is a powerful resource for field-sensing \cite{budker2007natphys,balasubramanian2008nature,wrachtrup2011natphys,dyakonov2014scirep} and quantum information \cite{kimble2008nature,blatt2014review} applications. Coherent population trapping (CPT) of spins \cite{fleischhauer2005rmp,Optical_control_Awschalom,Acosta_NV_multipass_PRL} is here fundamental to all-optical control.
When the interaction with a spin is weak, addressing an ensemble of identical spins can give a collectively-enhanced\cite{duan2001nature,lukin2003rmp,weimer2013prl}, strong interaction. For work with solids, favorable spin properties were identified for the nitrogen-vacancy defect in diamond \cite{Santori1,Santori2,hanson2008nature,dobrovitski2013arevcmp,Optical_control_Awschalom}
and divacancies in SiC \cite{janzen_deep_defects_shown,GaliTheory,koehl2011nature,falk2013natcomm,Strain_SiC_divacancies_Falk,christle2014arxiv,flipflop_protection_wrachtrup}.
However, for such defects inhomogeneities often impede resonant optical addressing of ensembles.
Besides inhomogeneity for the optical transition frequencies, defects show a distribution of orientations in the crystal. This prohibits homogenous interaction with fields since the orientation sets the direction of the electric dipole moment and crystal field for spin. The diatomic layering of SiC offers here an advantage over diamond in that it partly removes degeneracies for defects in different orientations \cite{koehl2011nature,falk2013natcomm}.
Here we demonstrate all-optical addressing and CPT for spin states of SiC divacancies, selectively on ensembles in particular crystal directions. CPT was realized after using two-laser control techniques for identifying the spin $S=1$ structure of the ground and optically excited states, and a role for relaxation pathways via intersystem crossing \cite{Baranov_V_SiC_identification_and_ISC_JETP}. Our results show that defect spin ensembles in SiC are promising systems for advancing the aforementioned applications, including the technically robust spin-ensemble approaches \cite{duan2001nature} to photon-mediated quantum networks with spins in solid state \cite{kimble2008nature}.
With the well-developed semiconductor processing for SiC \cite{SiCindustry,song2011optexpr} and the near-telecom value for the divacancy optical transition wavelength, research on integrated quantum device structures with existing technologies is within reach.

\begin{figure}[h!]
\centering
\includegraphics[width=8.9cm]{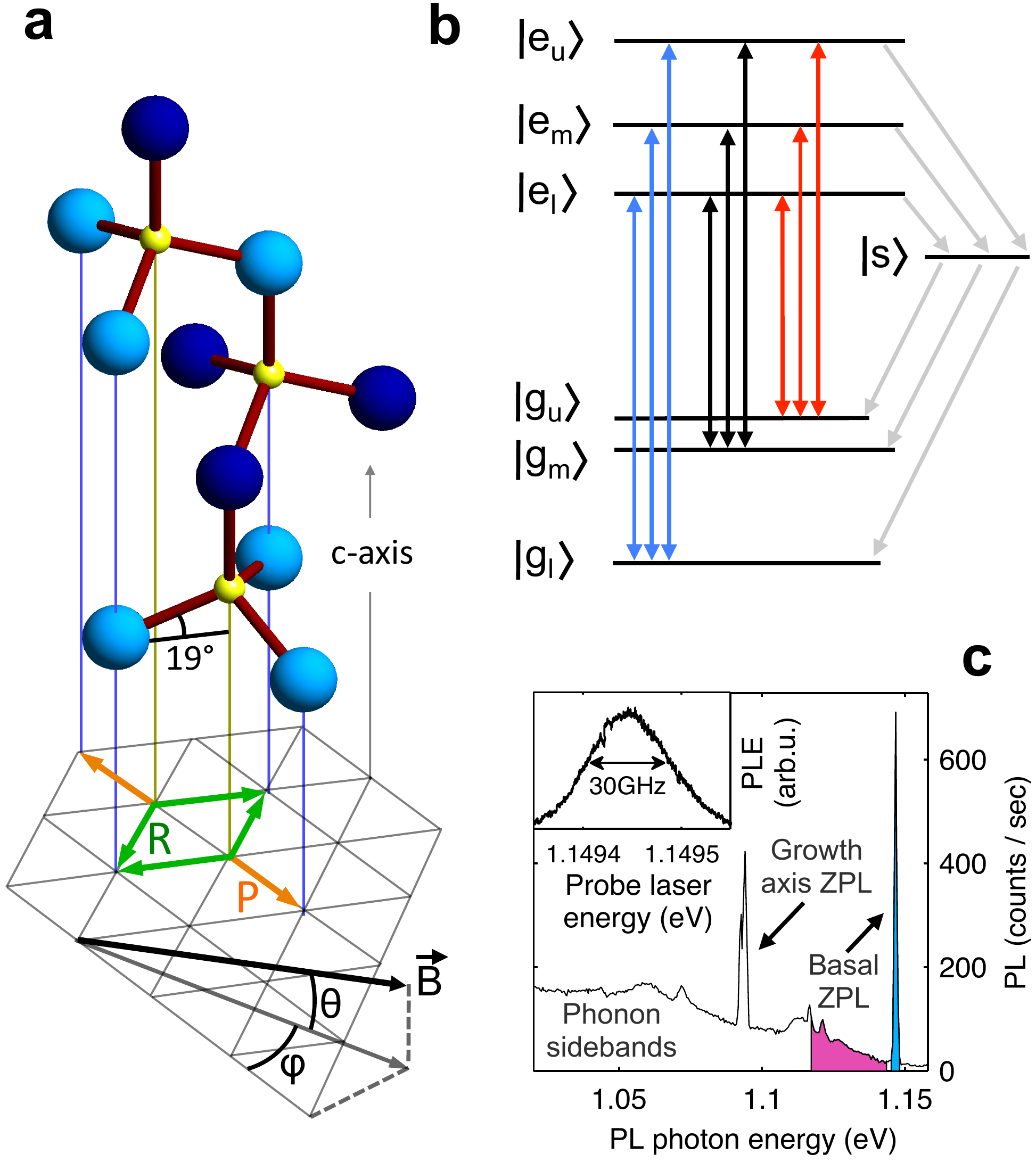}
\caption{\textbf{Crystal structure, energy levels, and optical signatures of divacancy defects in 4H-SiC.}
\textbf{a}, In the 4H-SiC crystal structure one can recognize carbon (yellow)-centered tetrahedrons with four silicon atoms (blue) at the corners. Dark and light blue signify silicon layers that have different crystal environments. This gives different optical transition energies for $V_{SiC}$ along the $c$-axis (vertical) and $V_{SiC}$ in the other six directions (basal-plane $V_{SiC}$, projections indicated on the bottom plane in orange, green, for sub-ensembles $P$, $R$, respectively). The magnetic field is applied as indicated, parallel to the orange $V_{SiC}$ projections (with small misalignment angles $\theta$, $\varphi$). Laser beams propagate near-parallel with the $c$-axis (Methods, Supplementary Information).
\textbf{b}, Level structure for the transitions of the zero-phonon line (ZPL) for basal $V_{SiC}$. The ground state and excited state both have a triplet $S=1$ spin structure (see main text for details). Vertical arrows indicate the nine possible optical transitions, where colouring labels the involved ground-state level $\ket{g_i}$. Gray lines indicate inter-system-crossing (ISC) relaxation pathways (via a singlet state $\ket{s}$).
\textbf{c}, Photoluminescence (PL) from 4H-SiC at 12~K, showing ZPL of $V_{SiC}$ in different crystal environments and their overlapping phonon sidebands (PSB). The blue-shaded ZPL (known as PL4) belongs to the basal $V_{SiC}$. In photoluminescence-excitation (PLE) studies lasers are resonant with PL4, while photons emitted in the PSB (pink) are used for detection. Inset: PLE spectrum of the PL4 line reveals an inhomogeneous linewidth of 30~GHz.}
\label{Fig:FigLatticePL}
\end{figure}

The divacancies (missing neighboring carbon and silicon atoms, here denoted as $V_{SiC}$) 
occur naturally in our high-purity, semi-insulating wafer, which was obtained commercially (Methods). Figure~\ref{Fig:FigLatticePL}a presents the lattice and possible $V_{SiC}$ orientations in the 4H polytype we work with. Basal-plane $V_{SiC}$ occur in six different directions, as indicated in the grid at the bottom. These have equivalent crystal environments and (for zero magnetic field) identical optical transition frequencies. The $V_{SiC}$ along the $c$-axis have different transition frequencies \cite{koehl2011nature,falk2013natcomm}, and are thus an obvious choice for addressing unidirectional ensembles. Our experiments focus nevertheless on basal $V_{SiC}$ since this allows for exploiting symmetries between $V_{SiC}$ in particular directions. We study how an external magnetic field can either define or break these symmetries, which is of interest for field-sensing applications. Specifically, we selectively address sub-ensembles of basal $V_{SiC}$ in particular directions by applying a weak magnetic field in the basal plane, in Fig.~\ref{Fig:FigLatticePL}a parallel to the two orange arrows (defining sub-ensemble $P$, Parallel), and hence at 60 degrees to the four green arrows (sub-ensemble $R$, Rotated). Small misalignment angles are labeled $\theta$ and $\varphi$. Due to the anisotropy of the spin $S=1$ Hamiltonian for $V_{SiC}$ the $P$ and $R$ sub-ensembles respond differently to the applied field. However, for zero misalignment, the symmetry within these sub-ensembles gives a homogenous response to the applied magnetic field and laser driving for both $P$ and $R$.

For the electronic ground (g) and excited (e) state of $V_{SiC}$, the spin Hamiltonian has the form \cite{GaliTheory,falk2013natcomm}
\begin{equation}
	H_{g(e)} = g_{g(e)}\mu_{B}\vec{B} \cdot \vec{S} + hD_{g(e)}S_{z}^{2} + hE_{g(e)}(S_{x}^{2} - S_{y}^{2}),
\label{Eq:HamilDE}
\end{equation}
where $g_{g(e)}$ is the g-factor, $\mu_{B}$ is the Bohr magneton, $\vec{B}$ is the applied magnetic field, $\vec{S}$ is the unitless spin $S=1$ operator, and $h$ is Planck's constant. $D_{g(e)}$ and $E_{g(e)}$ are the crystal-field splitting parameters in Hz, from spin-spin interaction and crystal anisotropy, respectively. The $z$-axis points along the divacancy axis from the missing Si-atom to missing C-atom position.

Figure~\ref{Fig:FigLatticePL}b illustrates the energy eigenstates $\ket{g_i}$ and $\ket{e_j}$ (with $i,j=l[ower],m[iddle],u[upper]$) of equation~(\ref{Eq:HamilDE}), which are superpositions of pure spin $S_{z}$ states. The nine double-headed arrows indicate the possible direct optical excitation and decay pathways, and we will use blue, black and red colouring for transitions that couple to $\ket{g_l}$, $\ket{g_m}$ and $\ket{g_u}$, respectively. Gray arrows indicate alternative non-radiative decay paths from levels $\ket{e_j}$ to $\ket{g_i}$ via a singlet state $\ket{s}$ \cite{Baranov_V_SiC_identification_and_ISC_JETP}. This process, known as intersystem crossing (ISC), also occurs for nitrogen-vacancy (NV$^-$) centers in diamond \cite{LukinISC} where it can yield high-fidelity spin initialization due to preferred relaxation into $\ket{g_l}$. Finally, divacancies can bleach under optical excitation (not depicted in Fig.~\ref{Fig:FigLatticePL}b). Here the divacancy alters its charge state and becomes off-resonant with the driving lasers (also found for NV$^-$ centers \cite{beha2012prl}).
Our measurements show laser-frequency selective bleaching that can last for hours, and that it can be rapidly reversed with a 685-nm repumping laser (Supplementary Information). Notably, such selective bleaching can be applied for
removing inhomogeneity for the optical transitions \cite{hole-burning-book} (but we do not apply this in our present study).

For initial characterization of our material we studied the photoluminescence spectrum from above-bandgap laser illumination (380~nm) at 12~K (Fig.~\ref{Fig:FigLatticePL}c). Several sharp zero-phonon lines (ZPL) from divacancy defects in different crystal environments come from the direct optical decay between levels as in Fig.~\ref{Fig:FigLatticePL}b. Each of these lines is accompanied by a broad phonon sideband, stretching to lower energies. The blue-shaded ZPL near 1.15~eV (known as PL4 \cite{koehl2011nature}) belongs to the basal $V_{SiC}$ we focus on. Next, we resonantly address an ensemble of these divacancies with excitation lasers tuned to this ZPL, and collect light emitted in the phonon sideband between 1.145 and 1.120~eV (purple-shaded part Fig.~\ref{Fig:FigLatticePL}c). When scanning a laser across this ZPL the phonon-sideband emission is proportional to the excitation into levels $\ket{e_j}$, and the resulting spectrum has a resolution set by the laser accuracy (1~MHz): a technique known as photoluminescence excitation (PLE). The inset in Fig.~\ref{Fig:FigLatticePL}c shows such a PLE spectrum from scanning a single laser across the PL4 line at 16~K, revealing a ZPL that has an inhomogeneous width of 30~GHz, which smears out the spectral fingerprint of particular $\ket{g_i}$-$\ket{e_j}$ transitions. We attribute this inhomogeneity to strain in the sample\cite{Strain_SiC_divacancies_Falk}.

\begin{figure}[h!]
\centering
\includegraphics[width=16cm]{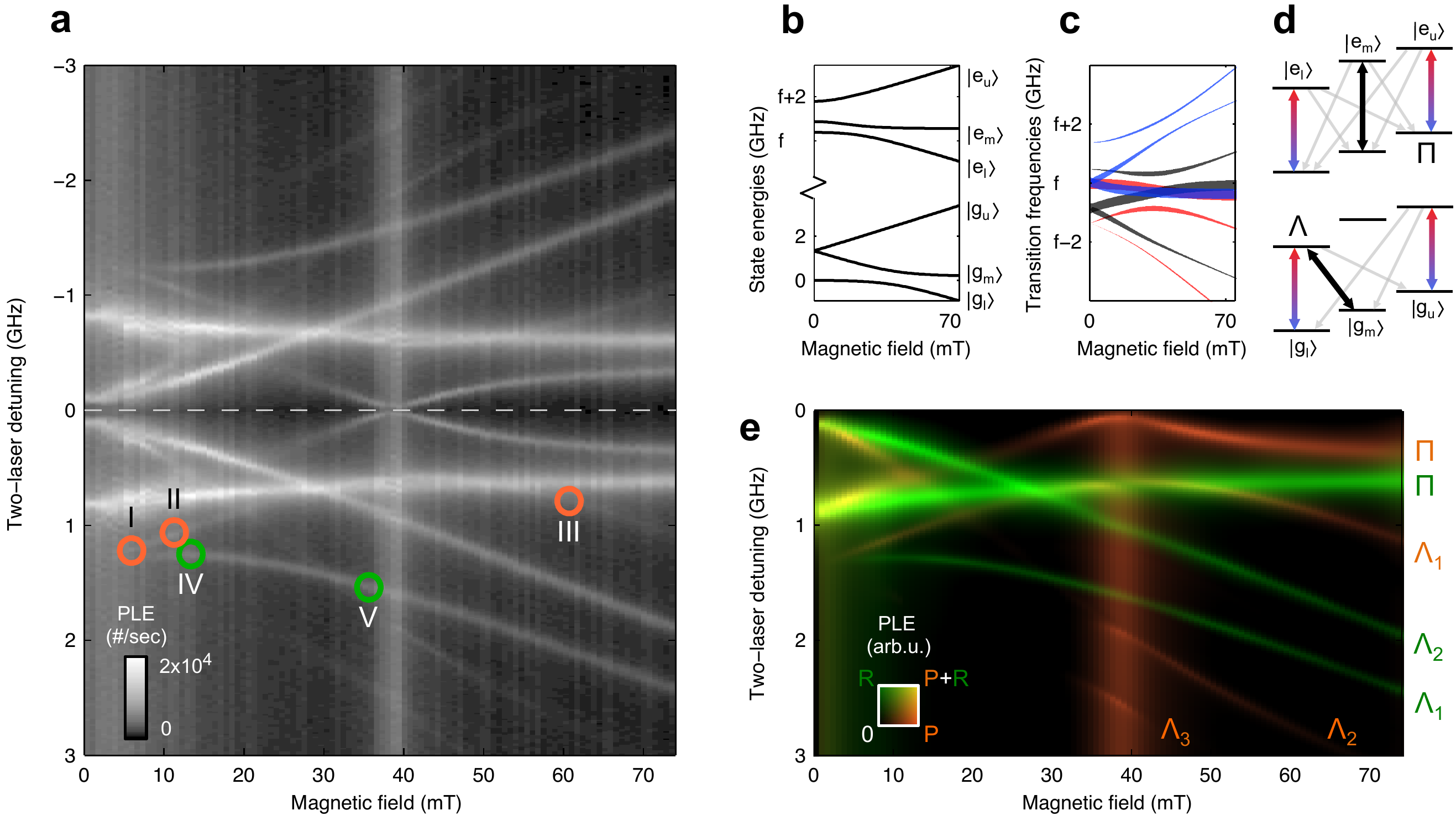}
\caption{\textbf{Two-laser magneto-spectroscopy of spin-related fine structure in the PL4 transition.}
\textbf{a}, PLE emission of the divacancies as a function of the frequency difference between the applied lasers, and applied magnetic field. The symmetry around 0~GHz detuning is inherent to the method (see Methods). Labels at coloured markers refer to Fig.~\ref{Fig:FigCPTtraces}.
\textbf{b}, Calculated energy levels of the ground and excited state spin $S=1$ systems, separated by transition frequency $f$, as a function of magnetic field (here for defect orientations $P$). The traces reflect the competition between the magnetic and crystal field terms in equation~(\ref{Eq:HamilDE}).
\textbf{c}, Based on panel \textbf{b}, the frequencies of the nine transitions of Fig.~\ref{Fig:FigLatticePL}b as a function of magnetic field (for orientations $P$). The colour-coding of Fig.~\ref{Fig:FigLatticePL}b is used again to specify the involved $\ket{g_i}$.
The width of traces represents the transition linewidth.
To get bright PLE, optical pumping into one of the three long-lived ground-state levels $\ket{g_i}$ has to be avoided by having lasers resonant with a red, black and blue transition. Where two transitions of different colour overlap (for example, the blue and red transitions near the central frequency $f$) this can be realized with two lasers.
\textbf{d}, Examples of two-laser pumping schemes where one laser (blue-red arrow) is resonant with two transitions (blue and red from panel \textbf{c}) while the other laser (black arrow) is resonant with one transition. Gray arrows are decay paths. Schemes where the lasers (do not) couple two different levels $\ket{g_i}$ to the same level $\ket{e_j}$ are termed $\Lambda$ ($\Pi$).
\textbf{e}, Calculated PLE levels from fitting our theoretical modeling to the data in panel \textbf{a}. Green and orange colour indicate emission from the $P$ and $R$ sub-ensembles. Lines are labeled according to their $\Lambda$ or $\Pi$ pumping scheme.}
\label{Fig:FigPLEspec}
\end{figure}

For investigating the spin-related fine structure within the ZPL, we use a two-laser spectroscopy technique \cite{manson1984jlum,Santori1} that
gives spectral features that are governed by the homogeneous optical linewidth.
It reveals PLE signals from a sub-ensemble (here not in the sense of $P$, $R$, but with respect to inhomogeneity for the transition) of $V_{SiC}$ with homogeneous transition frequencies.
We exploit that our system only gives high PLE signal when one laser is simultaneously resonant with transitions from two different $\ket{g_i}$ levels, while the other laser is resonant with a transition from the third $\ket{g_i}$ level (Methods).
At these fields and laser detunings optical pumping into one of the long-lived levels $\ket{g_i}$ is prevented, such that the PLE from this particular sub-ensemble does not darken. Magneto-spectroscopy results of such two-laser studies are presented in Fig.~\ref{Fig:FigPLEspec}a, obtained with one laser fixed and the other scanning near-central on the inhomogeneously broadened ZPL, and the sample at 4.2~K. The results show several bright PLE lines, which identify points where transition energies from two different levels $\ket{g_i}$ are identical to within their homogeneous linewidths. This occurs in particular for the levels $\ket{g_l}$ and $\ket{g_u}$, as illustrated in
Fig.~\ref{Fig:FigPLEspec}b-d. Figure~\ref{Fig:FigPLEspec}b shows the calculated evolution with magnetic field (using parameters derived from our measurements, see below) of the $\ket{g_i}$ and $\ket{e_j}$ levels. Figure~\ref{Fig:FigPLEspec}c shows the corresponding optical transition frequencies, where the width of the traces represents the transition linewidth (Supplementary Information). For many field values transitions that couple to $\ket{g_l}$ (blue) and $\ket{g_u}$ (red) have a near-identical frequency. This yields optical excitation schemes as in Fig.~\ref{Fig:FigPLEspec}d.

The curved PLE lines in Fig.~\ref{Fig:FigPLEspec}a allow for a detailed analysis of the parameters $D_{g(e)}$ and $E_{g(e)}$ of equation~(\ref{Eq:HamilDE}). This also yields detailed insight into the spin overlap $\braket{g_i|e_j}$, which governs the strength of a $\ket{g_i}$-$\ket{e_j}$ optical transition, and thereby the amplitude of the PLE signals (Franck-Condon principle with respect to spin\cite{fox2010book}). We fit the results of Fig.~\ref{Fig:FigPLEspec}a with a model (Supplementary Information) that combines rate equations for transitions with solving equation~(\ref{Eq:HamilDE}). The results are presented in Fig.~\ref{Fig:FigPLEspec}e, with green and orange shading representing PLE from the sub-ensembles in $P$ and $R$ orientations.

The fit closely resembles the data in Fig.~\ref{Fig:FigPLEspec}a in nearly all features. For example, the increase in PLE background around 38~mT results from a single laser pumping from all three ground states at that particular field for sub-ensemble $P$ (the point where transitions of three colours cross in the center of Fig.~\ref{Fig:FigPLEspec}c). With $D_g=1.334~{\rm GHz}$ and $E_g=18.7~{\rm MHz}$ from literature \cite{koehl2011nature} (consistent with our measurements), the fitting yields excited-state parameters $D_{e}=0.95\pm0.02$~GHz and $E_{e}=0.48\pm0.01$~GHz, and a rate $G_0 = 20\pm 5$~MHz for the radiative contribution to the homogeneous linewidths. Getting detailed agreement between the modeling and the data required inclusion of intersystem crossing rates between 1 and 14~MHz, with a dependence on magnetic field (Supplementary Information). The analysis for Fig.~\ref{Fig:FigPLEspec}e also identifies for each PLE line whether the underlying pumping scheme is of the $\Pi$ or $\Lambda$ type (defined in Fig.~\ref{Fig:FigPLEspec}d).

We next show that our two-laser addressing of systems with a three-level ground state is suited for coherent control of the spin states. Of particular relevance is coherent population trapping (CPT), a key effect in quantum-optical control of spins \cite{fleischhauer2005rmp}. Here, two-laser driving of two states $\ket{g_i}$ to a common state $\ket{e_j}$ shows --on exact two-photon resonance-- destructive quantum interference in the dynamics to the excited state, which results in coherent control of the ground state. Specifically, for the $\Lambda$ driving scheme as in Fig.~\ref{Fig:FigPLEspec}d, the system would (in the case of ideal spin coherence) get trapped in the state
\[\ket{\Psi_{CPT}} \propto \Omega_{m}\ket{g_l} -  \Omega_{l} \ket{g_m} + 0\ket{g_u},\]
where $\Omega_{l}$ and $\Omega_{m}$ are the Rabi frequencies for the driven transitions from $\ket{g_l}$ (blue-red arrow) and $\ket{g_m}$ (black arrow), respectively. Notably, this example uses again that the doubly-resonant laser avoids population trapping in $\ket{g_u}$. This can be directly applied in schemes where the blue-red arrow in Fig.~\ref{Fig:FigPLEspec}d is a control field and the black arrow a signal field.

\begin{figure}[h!]
\centering
\includegraphics[width=10cm]{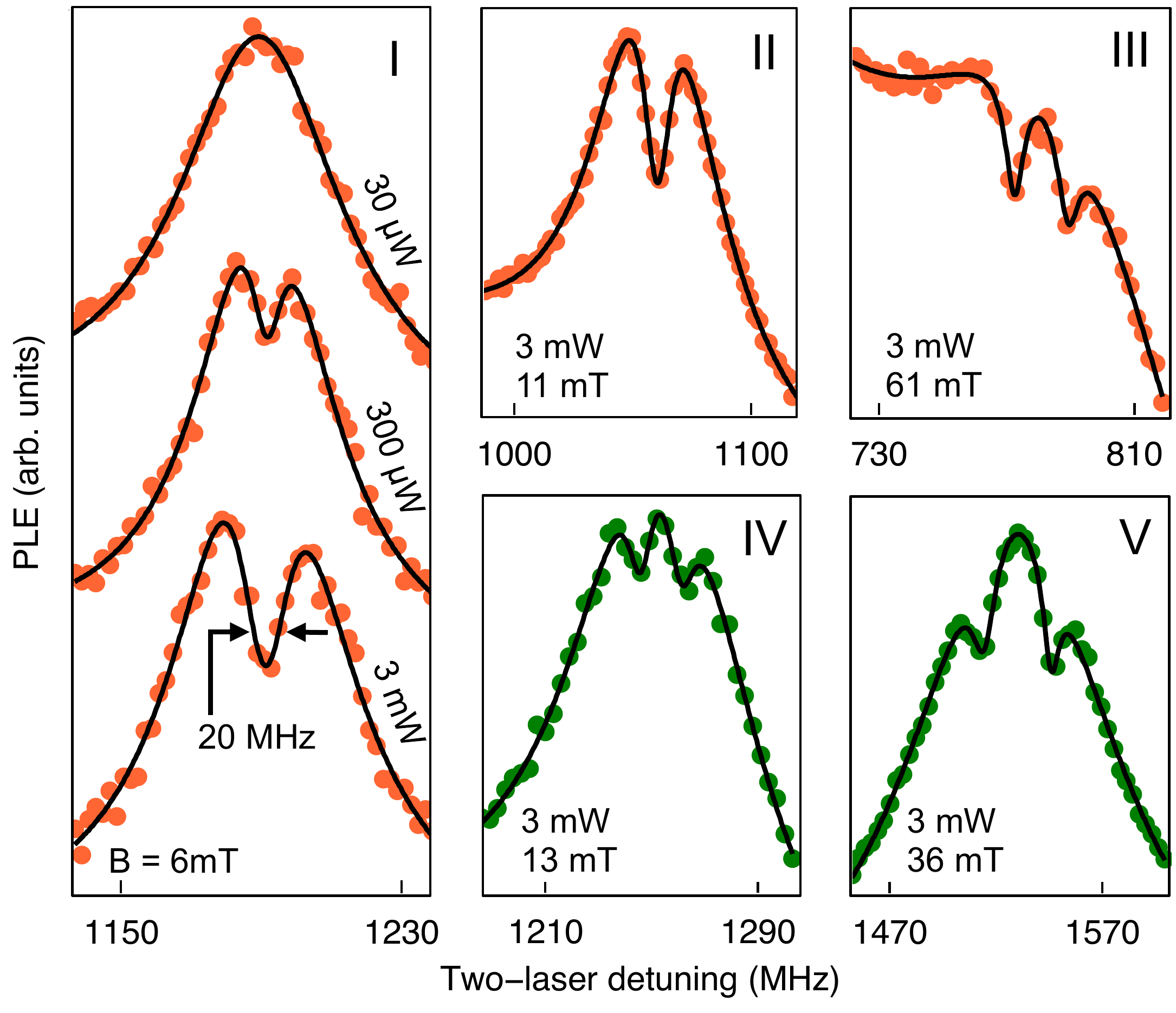}
\caption{\textbf{Spectral signatures of coherent population trapping.}
Panels I through V show dips from coherent population trapping (CPT) in PLE spectra taken at locations I-V in Fig.~\ref{Fig:FigPLEspec}a (applied laser powers and magnetic fields as indicated). Panel~I shows the emergence a CPT dip with increasing laser power, in the two-laser PLE lines (offset for clarity) for the $\Lambda_1$ line of defect orientations $P$ (orange). Panels II through V present CPT dips in the $\Lambda_{1}$-lines for both $P$ and $R$ orientations at increasing magnetic field, showing a gradual splitting of the CPT feature. This reflects small misalignments with magnetic field $\theta=0.8^\circ$ and $\varphi=1.8^\circ$ (as defined in Fig.~\ref{Fig:FigLatticePL}a). Solid lines are fits for a theoretical model of CPT.}
\label{Fig:FigCPTtraces}
\end{figure}

For studying the occurrence of CPT, we focus on the PLE lines labeled $\Lambda_{1}$ in Fig.~\ref{Fig:FigPLEspec}a. Figure~\ref{Fig:FigCPTtraces} presents PLE spectra taken at the locations marked I through V in Fig.~\ref{Fig:FigPLEspec}a, with both data for orientations $P$ (orange markers) and $R$ (green markers).
Panel I shows how a central dip appears in the PLE line for sub-ensemble $P$ as the laser power is increased. This is the spectral signature of CPT, where trapping in a ground-state superposition causes a quenching of the optical excitation.
For our measurements the amplitude of the dip in the PLE signal is suppressed due to our experimental geometry, which
gives decaying intensities for the lasers fields while they propagate in the sample (Supplementary Information).

For the observed CPT dips we find good agreement with calculations (solid lines in Fig.~\ref{Fig:FigCPTtraces}, Supplementary Information). This yields a ground-state dephasing time of 42$\pm$8~ns, which is about 30 times shorter than previously reported from electron-spin-resonance studies on a comparable sample \cite{koehl2011nature}.
Such a discrepancy is likely due to extra dephasing arising from the permanent strong laser driving, which causes a fluctuating Stark effect\cite{Strain_SiC_divacancies_Falk} from charges that move in and out of localized traps, as also reported for NV$^-$ centers\cite{Acosta_NV_multipass_PRL}.
This can be avoided by using pulsed laser control or reduced powers, which is inherent to the envisioned quantum applications of $V_{SiC}$ \cite{Optical_control_Awschalom,Acosta_NV_multipass_PRL}.

In panels II through V, the evolution of the CPT dip with magnetic field is shown for both the $P$ and $R$ sub-ensembles for constant laser powers of 3~mW. The CPT dip clearly splits at higher magnetic fields. This is caused by small misalignment angles of $\theta = 0.8\pm0.2^\circ$ and $\varphi = 1.8\pm0.2^\circ$ that break the symmetry between the two (four) different defect orientations within the $P$ ($R$) sub-ensemble.
We confirmed this by tuning the misalignment angles (Supplementary Information).
Due to their different symmetries, splittings in $P$ are particularly sensitive to variations in $\theta$, while splittings in $R$ respond predominantly to variations in $\varphi$.
For optimized sample geometries, and with the reported spin coherence times \cite{koehl2011nature,christle2014arxiv}, the ultimate accuracy for probing the CPT-dip frequencies should be at the kHz level\cite{fleischhauer2005rmp}. Sensitivity to field alignment varies between 15 and 25~MHz per degree for different CPT splittings at 70~mT (Supplementary Information), which gradually decreases for lower fields. Here the CPT sensitivity to field magnitude is $\sim$30~MHz/mT. Our results thus identify the potential for unidirectional divacancy ensembles for CPT-based field-sensing and quantum information applications.

\newpage

\noindent \textbf{METHODS SUMMARY}\\
We studied a 4H-SiC sample cleaved from a 365-$\mu$m thick wafer (purchased at CREE Electronics, product HPSI W4TRD0R-0200). Its shape and a partial gold coating were designed for enhancing the amount of divacancy emission while geometrically separating the weak divacancy emission from control laser beams (Supplementary Information). The sample was mounted in a flow cryostat with windows for free-space optics. Two tunable continuous-wave diode lasers were used for two-laser spectroscopy near 1078.6~nm wavelength, with both lasers supplying equal power. The lasers had a linewidth below 1~MHz, frequencies that were stable to within 3~MHz over 10 minutes, while the two-laser detuning could be identified with 1~MHz resolution. An additional 685~nm continuous-wave diode laser was used to counter the bleaching of the divacancies. The diameter of the overlapping laser beams upon entering the sample was 80~$\mu$m. Divacancy emission was collected with a 1.2-cm focal-length lens, and sent to either a spectrometer for measuring PL, or to an InGaAs single-photon counter for measuring PLE. The applied magnetic field was provided by a superconducting magnet.

\vspace{0.5cm}

\noindent \textbf{METHODS}\\
\noindent \textbf{Sample fabrication}\\
The sample was cut from a 365-$\mu$m thick wafer, purchased at CREE Electronics, product HPSI W4TRD0R-0200 (http://www.cree.com/). The sample geometry was optimized for PLE detection (Supplementary Information). To cleave it into its precise final shape, a diamond-tipped stylus made a shallow groove in the wafer, after which it was broken along the groove by manually applying force with tweezers. A 100-nm gold mirror coating was evaporated onto the front and back of the sample, with a 1-nm layer of titanium for adhesion. We used a hard mask to keep a small region free for the lasers to couple into the sample.

\noindent \textbf{Experimental setup}\\
The sample was placed in a liquid-helium flow cryostat, where the sample temperature was kept stable to within 0.01~K with a PID controller managing the helium flow and a sample heater. Windows on the four sides of the square cryostat allowed for optical access to the sample. Two tunable, continuous-wave diode lasers provided excitation light near the ZPL wavelength 1078.6~nm. We present here results taken with equal powers for the two lasers. The lasers had a linewidth below 1~MHz, and were stable to within 3~MHz over 10 minutes. This was characterized by having the two lasers interfere on a photodetector, and monitoring the shape of the beat envelope, and time-evolution of the beat frequency, caused by a minor detuning of the lasers. The two lasers beams entered the sample with identical linear polarization, and we did not observe a dependence on changing this polarization. Laser powers were varied between 30~$\mu$W and 3~mW with neutral density filters. To counter bleaching, we added a 685~nm diode laser which was permanently on.

The frequencies of the two tunable lasers were monitored using a wavelength meter, containing several staged interferometers, which could be applied for getting sub-MHz frequency determination. The sample was mounted facing a window where excitation light entered, while the detected PL(E) emission left the sample at 90$^\circ$ from the excitation lasers. This light was collected by a 1.2-cm focal length lens mounted inside the cryostat. Outside, it was sent through additional long-pass filters with a 1082-nm cut-off to remove remaining light from the excitation lasers, and focussed into an 800-$\mu$m core multimode fiber. This was connected to a spectrometer to measure PL, or to an InGaAs single photon counter (range 900 to 1150~nm) for PLE. The magnetic field was generated by a superconducting magnet, and was fixed along the direction of PL collection, in the plane of the sample.

\noindent \textbf{Measurement techniques}\\
For the two-laser spectroscopy data of Fig.~\ref{Fig:FigPLEspec}, one laser was fixed central on the ZPL. The second laser was used for frequency scanning around the frequency of the fixed laser at a rate of 10~MHz/sec. For an ensemble with inhomogeneous transition frequencies, the roles of the two lasers must nevertheless be equivalent, and this underlies the symmetry around 0~GHz detuning in Fig.~\ref{Fig:FigPLEspec}a.
Frequencies of both lasers were recorded every few milliseconds, to allow for accurate frequency binning of photons counts, and quick automated correction for frequency mode hops of the lasers. The accumulated number of counts from the photon counter were read out from a time-to-digital converter once per second. Together with the scan rate this set the (minimum) binning resolution to 10~MHz. This was then repeated for different magnetic fields. At the start of each scan, the PLE from the fixed laser was measured, and was subtracted from the data. For the traces in Fig.~\ref{Fig:FigCPTtraces}, the scanning laser was repeatedly scanned around the frequency of the fixed laser over a range of a few hundred MHz, with 7~s scan duration. To obtain a higher frequency resolution than in the experiment for Fig.~\ref{Fig:FigPLEspec}, the arrival time of each photon was correlated with the frequency measurement times, allowing for a resolution well below 1~MHz.

\newpage

\noindent \textbf{REFERENCES FOR THE MAIN TEXT}

\vspace{1cm}

\noindent \textbf{Supplementary Information} is included as an appendix to this document.

\vspace{1cm}

\noindent \textbf{Acknowledgements}\\
We thank J.~P.~de Jong, D.~Perdok, A.~Singh, X.~Yang, X.~Bonet Monroig, M. de Roosz, R.~W.~A.~Havenith, A.~Alkauskas, W.~Gomulya and M.~A. Loi for help and discussions. Financial support was provided by ERC Starting Grant 279931.

\vspace{1cm}

\noindent \textbf{Author Contributions} C.H.W. and A.R.O initiated the project. O.V.Z. was the lead researcher for experiments, data analysis, and writing the paper. All authors contributed to experiments, data analysis and improving the paper.

\newpage


\newpage



\newpage


\include{Zwier-SiC-CPT-SI}

\end {document}

%% file: Zwier-SiC-CPT-SI.tex
\setcounter{figure}{0}

\renewcommand{\thesection}{\arabic{section}}
\renewcommand{\thefigure}{S\arabic{figure}}

\makeatletter
 \renewcommand\@biblabel[1]{#1.}
\makeatother


\newcommand{\be}[1]{\begin{eqnarray}  {\label{#1}}}
\newcommand{\ee}{\end{eqnarray}}

\begin{center}
\textbf{{\LARGE Supplementary Information}}

for

\textbf{{\Large All-optical coherent population trapping with\\ defect spin ensembles in silicon carbide}}

by

Olger~V.~Zwier, Danny~O'Shea, Alexander~R.~Onur, and Caspar~H.~van~der~Wal

\end{center}

\vspace{2cm}


\noindent \textbf{TABLE OF CONTENTS}

\begin{figure}[h!]
\includegraphics[width=\columnwidth]{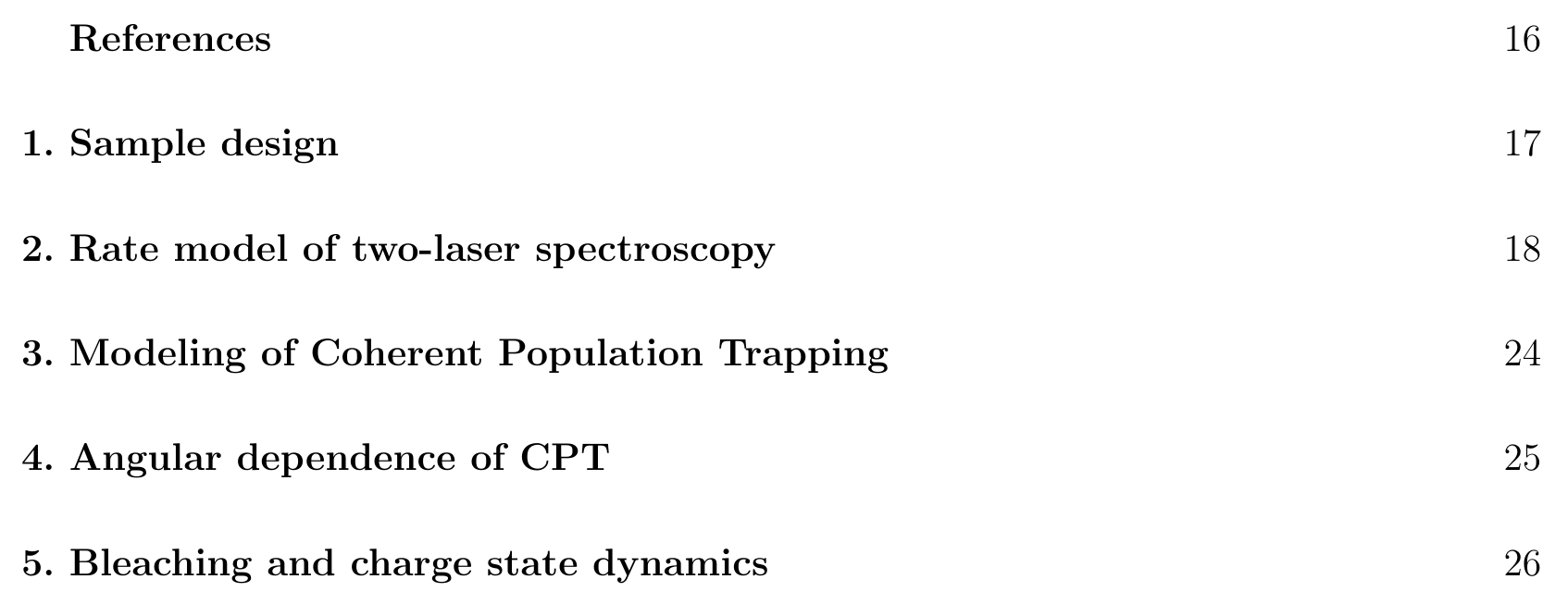}
\end{figure}


\newpage


\noindent \textbf{REFERENCES FOR THE SUPPLEMENTARY INFORMATION}

\newpage

\section{Sample design}
\label{Sec:SOMsampledesign}

\begin{figure}[t!]
\centering
\includegraphics[width=12cm]{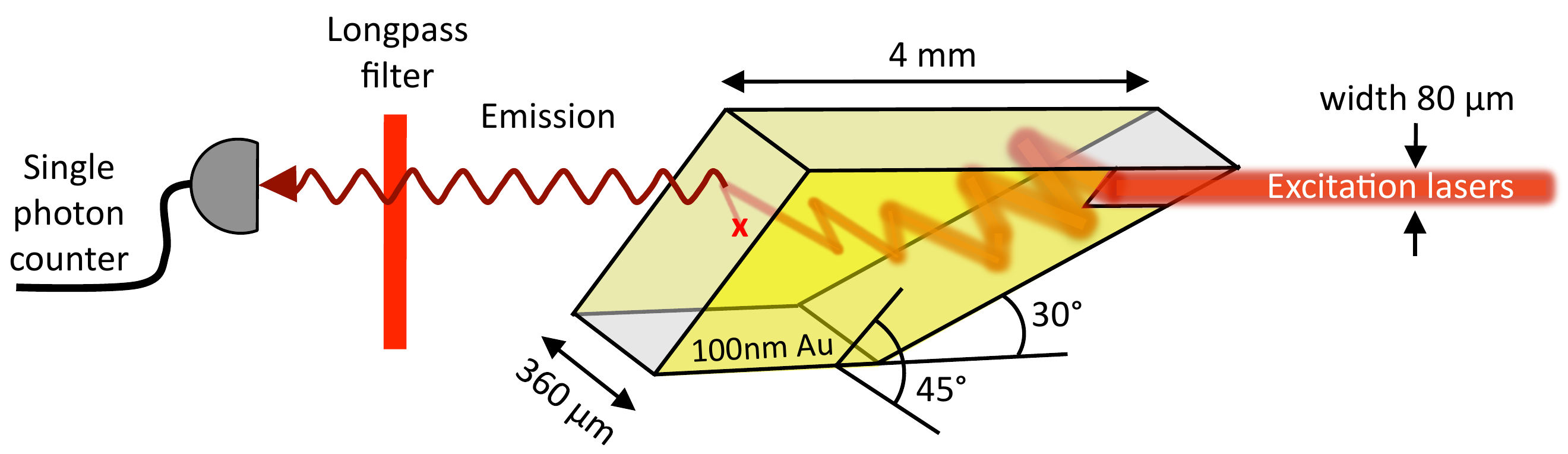}
\caption{\textbf{Multipass SiC sample.}
Schematic of the sample geometry used in our experiment. Lasers are incident on the sample from the right, entering through a window of 0.3~mm by 0.2~mm. Light propagates as shown, hardly exiting the sample at the marker x due to total internal reflection. This results in a factor $10^6$ suppression for laser light reaching PL(E) signal detectors. Light emitted by the divacancies does exit the sample, on the left. After additional spectral filtering it is collected by a single photon counter (or spectrometer). The trapezoidal shape of the sample, combined with gold coating on the front and back, further enhances this collection efficiency.}
\label{Fig:Fig_sample}
\end{figure}

We designed a sample structure that gave a high amount of PL and PLE emission, optimized PL(E) collection efficiency, while also geometrically separating the weak divacancy emission from the much stronger control laser beams (Fig.~\ref{Fig:Fig_sample}). Gold coatings of 100~nm thickness on the front and back of the sample act as mirrors that trap the applied laser beams in the sample. This enhances the path length of the laser in the sample, giving a higher number of divacancy defects that contribute to the emission signal. The laser beams enter the sample at the top-right corner (window of 0.3~mm by 0.2~mm without gold coating), while the sample has a 45$^\circ$ forward tilt with respect to the laser beam. Due to refraction at the interface this gives propagation in the sample that is near-parallel to the $c$-axis, with a deviation that is just large enough for trapping the light on a non-overlapping zig-zag trajectory. We measured a 10\% intensity loss at each reflection. In this manner laser light propagates until it reaches the left facet of the sample (no gold coating), denoted by an x, where most of the laser light still does not leave the sample due to total internal reflection. A small fraction does exit because the interfaces are not perfectly smooth. This approach give a suppression of $10^6$ in laser light reaching our detectors. Conversely, divacancy emission is in random directions and for a large part on trajectories that do leave the sample when reaching the left facet. This includes trajectories that first get reflected inside the sample, and this is enhanced by total internal reflection at the trapezoidal shape on the right of the sample. This gives PL and PLE emission that mainly exits the sample on the left, in a cone of directions, facilitating efficient detection. We apply additional spectral filtering before our detectors for reaching an overall suppression of over $10^{10}$ for light from the excitation lasers.


\newpage

\section{Rate model of two-laser spectroscopy}
\label{Sec:SOMratemodel}

\begin{figure}[t!]
\centering
\includegraphics[width=6cm]{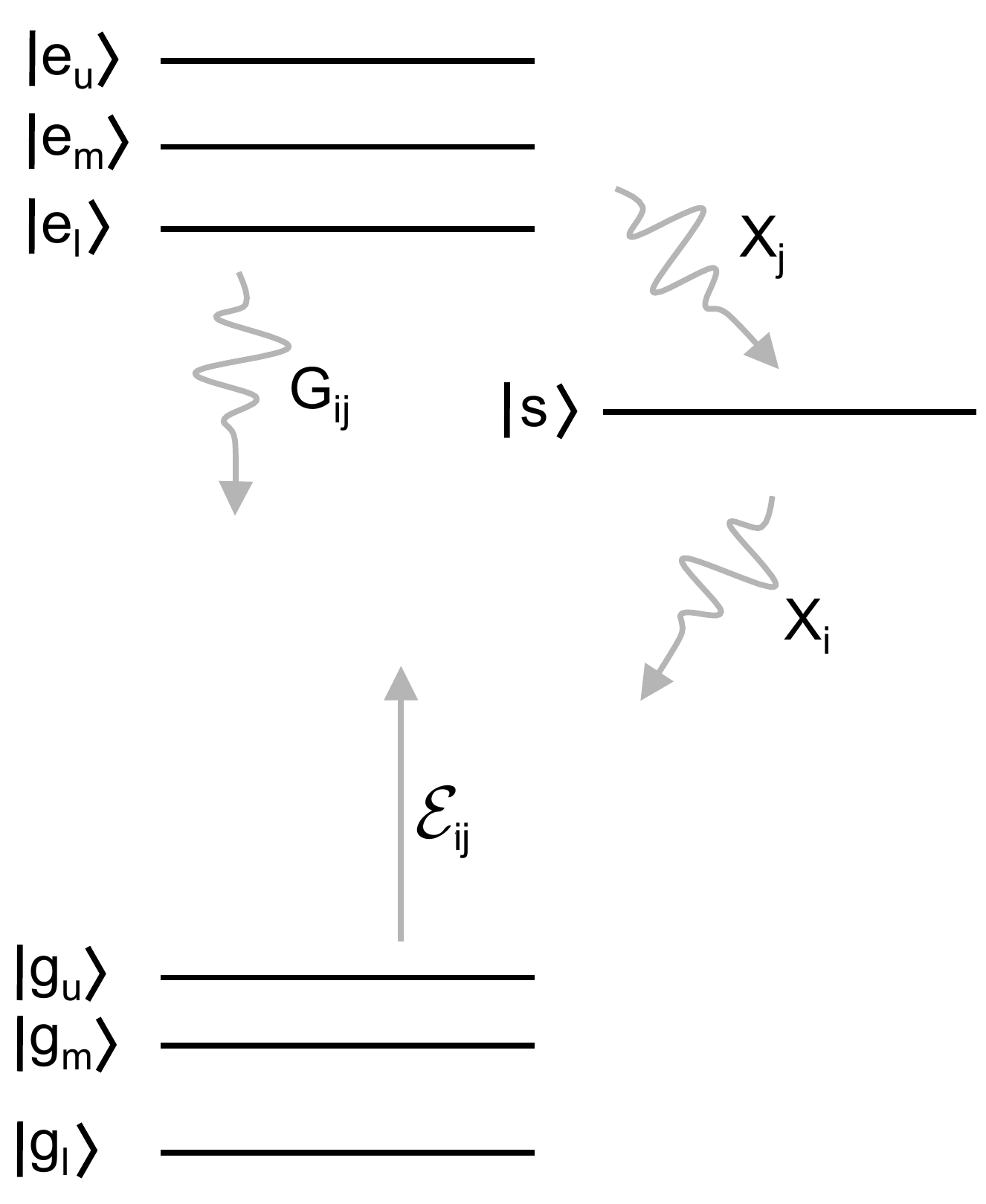}
\caption{\textbf{Schematic of energy levels and processes of the rate-equation model.}
$\mathcal{E}_{ij}$ is the rate of laser excitation from $\ket{g_i}$ to $\ket{e_j}$, $G_{ij}$ is the radiative decay rate from $\ket{e_j}$ to $\ket{g_i}$, $X^e_j$ is the intersystem-crossing (ISC) decay rate from $\ket{e_j}$ to $\ket{s}$, and $X^g_i$ is the ISC decay rate from $\ket{s}$ to $\ket{g_i}$.}
\label{Fig:Fig_rate_model}
\end{figure}

We model the spectral positions and amplitudes of PLE signals for data as in Fig.~\ref{Fig:FigPLEspec}a in an approach that combines rate equations for the transitions between the levels $\ket{g_i}$ and $\ket{e_j}$ with solving the spin Hamiltonian (equation~(\ref{Eq:HamilDE})) for the ground and excited state. After fitting this yields results as in Fig.~\ref{Fig:FigPLEspec}e.
The rate equations are the set of coupled equations for the processes and parameters that are illustrated in Fig.~\ref{Fig:Fig_rate_model}, and this system's steady-state solution determines the PLE signals from our experiments with continuous excitation.
We use an approach that only describes populations on the levels (and no coherences between the levels) in order to keep the total number of fitting parameters low. We repeated our experiments with variation of laser intensities over two orders of magnitude, and obtain nominally the same values for the fitting parameters (besides the trivial increase in PLE signal). This justifies that we can neglect coherences in our modeling.


The parameters for the rate equations, and the parameters $D_e$ and $E_e$ of the excited-state spin Hamiltonian, are obtained from fitting the model to data sets as in Fig.~\ref{Fig:FigPLEspec}a. We find parameters $D_g$ and $E_g$ for the ground-state spin Hamiltonian that are consistent with literature, and we therefore use these as fixed values. For a given magnetic field, the energy splittings between the three levels $\ket{g_i}$ and $\ket{e_j}$, and the associated spin states, are obtained by calculating eigenvalues and eigenvectors of equation~(\ref{Eq:HamilDE}) of the main text. The singlet state $\ket{s}$ is fixed. We calculate in parallel PLE signals from the $P$ and $R$ sub-ensembles, but for displaying the result in Fig.~\ref{Fig:FigPLEspec}e we use orange and green coloring for the respective signal contributions. For the small offset angles $\theta$ and $\varphi$ we had in our experiments (Fig.~\ref{Fig:FigLatticePL}a) the differences in eigenvalues for different divacancy orientations within the sub-ensembles $P$ and $R$ are negligible for the present analysis.

We start with describing the influence of the spin states on the radiative decay from population in a level $\ket{e_j}$ to a level $\ket{g_i}$. We first analyze this for the case of zero-phonon-line (ZPL) transitions (below we will further discuss the role of radiative decay via phonon-sideband emission), which occurs at a rate $G^z_{ij}$ and is given by
	\[G^z_{ij} = C_1 \, d_{ij}^2,\]
where $C_1$ is a constant and $d_{ij}$ is the transition dipole matrix element. We can write this out as
	\[G^z_{ij} = C_1|\bra{\psi^g_i}\hat{d}\ket{\psi^e_j}|^2,\]
where $\hat{d}$ is the electric dipole operator, and the states $\ket{\psi^e_j}$ and $\ket{\psi^g_i}$ are the initial and final state for this transition, respectively.
We will use the approximation that $\ket{\psi_i}$ can be treated as a product state of the spin state $\ket{g_i}$ and the orbital state $\ket{\chi^g_i}$ of the ground state:  $\ket{\psi^g_i} = \ket{\chi^g_i} \ket{g_i}$.
Similarly, we will use $\ket{\psi^e_j} = \ket{\chi^e_j} \ket{e_j}$ where $\ket{\chi^e_j}$ is the orbital state for the excited state. This approximation is valid for the case of very low spin-orbit interaction, and applies for our system given the low atomic masses of SiC.
We thus also assume that the orbital state $\ket{\chi^g_i}$ is identical for the three levels $\ket{g_i}$, and that $\ket{\chi^e_j}$ is identical for the three levels $\ket{e_j}$. Further, we assume that at the magnetic fields of our experiments the highly confined orbital states $\ket{\chi^g_i}$ and $\ket{\chi^e_j}$ have negligible magnetic field dependence.

Since the dipole operator only works on the orbital part of a state, the expression for $G^z_{ij}$ can now be written as
 	 \[G^z_{ij} = C_1|\bra{\chi^g_{i}}\hat{d}\ket{\chi^e_{j}}|^2~|\braket{g_i|e_j}|^2,\]
and with the above assumptions $|\bra{\chi^g_{i}}\hat{d}\ket{\chi^e_{j}}|$ is equal for all nine $\ket{e_j}$-$\ket{g_i}$ transitions. We can thus express $G^z_{ij}$ as
	\[G^z_{ij} = G^z_{0}|\braket{g_i|e_j}|^2.\]
Here $G^z_{0}$ is the total decay rate via ZPL emission out of the excited state, and $|\braket{g_i|e_j}|^2$ is the Franck-Condon factor with respect to spin\cite{SI-fox2010book}.

We need to account for the fact that a significant part of the radiative decay occurs via phonon-sideband emission. The ODMR work on the divacancies in SiC \cite{SI-koehl2011nature,SI-falk2013natcomm} shows that after phonon-sideband emission the spin state is preserved during subsequent phonon relaxation (which occurs at a fast time scale). Also, with the above assumptions, the rate of phonon-sideband emission is simply proportional to the ZPL emission for all $\ket{e_j}$-$\ket{g_i}$ transitions. We can thus simply express the total decay rate from all the possible radiative transitions as
rates $G_{ij}$ (we drop the superscript $z$) that obey
	\[G_{ij} = G_{0}|\braket{g_i|e_j}|^2, \]
and the PLE signal for a particular $\ket{e_j}$-$\ket{g_i}$ transition is proportional with $G_{ij}$ and the steady-state population in $\ket{e_j}$.
We treat $G_0$ as a fit parameter.

In addition to this radiative decay, we include parallel, non-radiative decay paths via the singlet state $\ket{s}$ for modeling intersystem-crossing (ISC) events. Each $\ket{e_j}$ has its own transition rate $X^e_j$ into $\ket{s}$, and from there the spin decays with rates $X^g_i$ to $\ket{g_i}$. We cannot exclude that the ISC rates depend on the spin state, and hence on magnetic field. We investigated fitting both with constant (field-independent) and field-dependent ISC rates. In the latter case, we treat all six ISC rates $X^g_{i}$, $X^e_{j}$ as fitting parameters which may slowly vary with magnetic field (further discussed below). During the fitting, we apply (in a self-consistent manner) that the homogeneous linewidth for the optical transitions is governed by the total decay rate $\Gamma_j$ out of a level $\ket{e_j}$, which we calculate as
$\Gamma_j = G_0 + X^e_j$.

The optical excitation rates $\mathcal{E}_{ij}$ are modeled in a manner similar to $G^z_{ij}$. Here we only need to consider ZPL transitions since the phonon-sideband excitation is negligible for laser excitation near resonance with the ZPL at the temperatures of our experiments. For excitation from a level $\ket{g_i}$ to a level $\ket{e_j}$
	\[\mathcal{E}_{ij} = C_2 \, d_{ij}^2 \, \frac{1}{(2\pi(f-f_{ij}))^2+\Gamma^2_j}  , \]
where $C_2$ is a constant, $f$ the laser frequency and $f_{ij}$ the resonance frequency of the transition. The last factor in this equation describes a Lorentzian lineshape of half-width $\Gamma_j/2\pi$ for the resonance (we checked that all our experiments were in the regime where the laser intensities give negligible power broadening of the PLE lines). We implement this with a fit parameter $\mathcal{E}_0$,
	\[\mathcal{E}_{ij} = \mathcal{E}_0 |\braket{g_i|e_j}|^2  \frac{1}{(2\pi(f-f_{ij}))^2+\Gamma^2_j}   .\]

The relaxation rates between ground state levels $\ket{g_i}$ are assumed to be negligible compared to the radiative decay and excitation rates (we checked that including these as parameters gave very low rates that do not improve the fits). For our fitting we subtract from the data a constant PLE background that is present from single laser driving of the inhomogeneous ensemble.

\vspace{1cm}

\begin{figure}[t!]
\centering
\includegraphics[width=7cm]{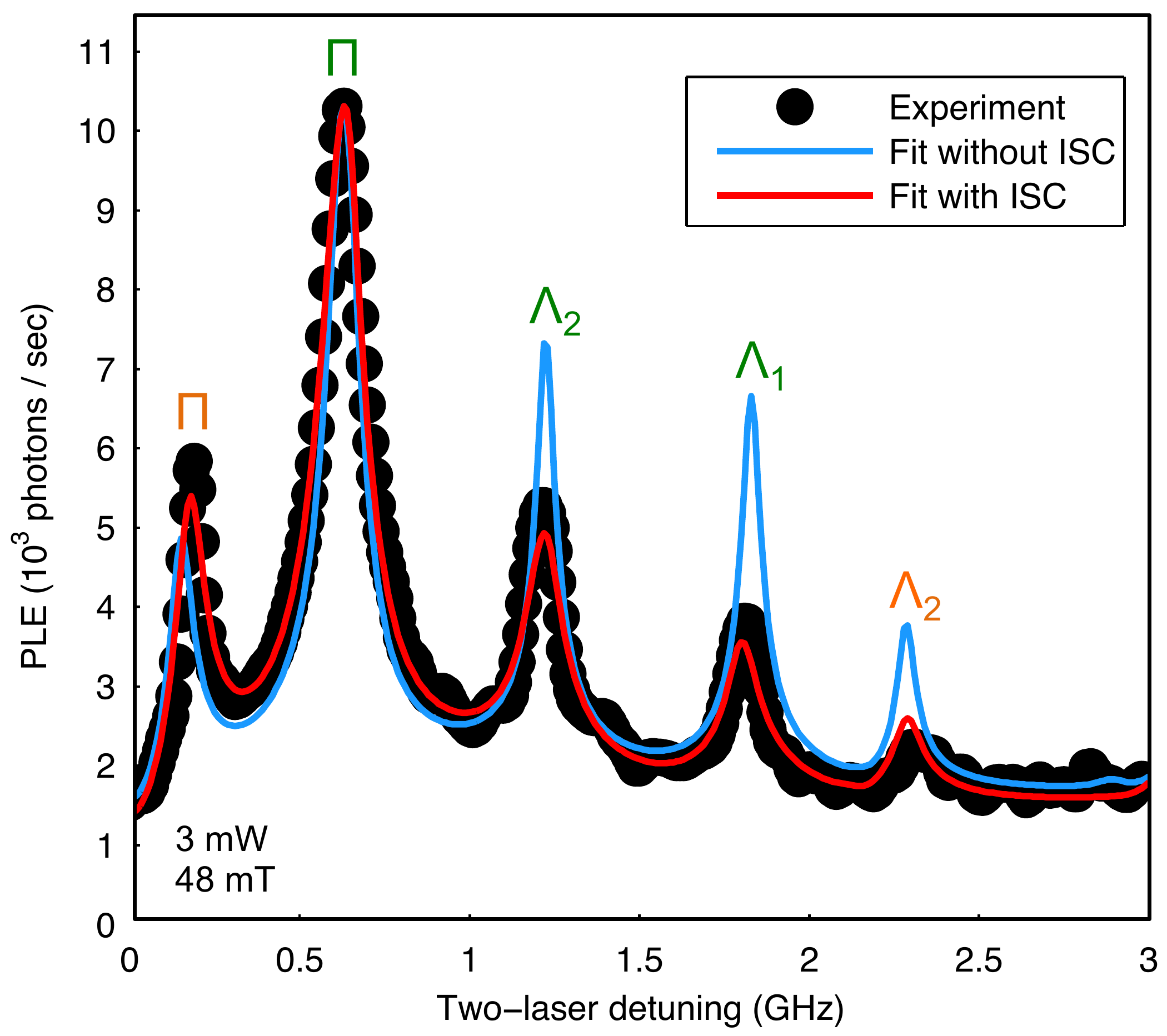}
\caption{\textbf{Model fits at 55~mT.}
Comparison of the quality of fitting the model, with (red) and without (blue) ISC. Without ISC, the model structurally overestimates peaks attributed to Lambda schemes.}
\label{Fig:Fig_single_fits}
\end{figure}

\noindent   \textbf{Fitting results and discussion}\\
\noindent The fitting yields $G_0=20 \pm 5~{\rm MHz}$, $D_e = 0.95\pm 0.02~{\rm GHz}$, and $E_e=0.48\pm 0.01~{\rm GHz}$ (and we used literature values \cite{SI-falk2013natcomm} $D_g = 1.334~{\rm GHz}$ and $E_g = 18.7$~{\rm MHz}).
The value for $\mathcal{E}_0$ is always such that it gives rates $\mathcal{E}_{ij}$ well below $G_0$ and shows no interdependence with the other parameters.
For these parameters only $G_0$ shows a weak dependence on the various approaches to fitting the ISC rates.
Attempts to fit without accounting for ISC yields PLE lines at the correct frequencies (mainly governed by the values of the $D_{e,g}$ and $E_{e,g}$ parameters) and with the correct width (mainly governed by $G_0$) in the plane of Fig.~\ref{Fig:FigPLEspec}a. However, for that case the relative amplitudes of calculated PLE lines (at a certain magnetic field value) show poor agreement with the experimental results. On this aspect we get much better agreement in an approach with magnetic-field-independent values for the six ISC rates (Fig.~\ref{Fig:Fig_single_fits}).
However, we cannot exclude that the ISC rates depend on magnetic field. While the ISC processes are far from completely understood, recent theory work for diamond NV$^-$ centers proposes a role for spin-orbit coupling, where the orbital quantization axis is linked to the defect orientation. Consequently, this work proposes that the ISC rates depend on the spin states of the levels, and thereby on magnetic field \cite{SI-LukinISC}. This also indicates that for non-zero magnetic field one should in fact fit with different ISC parameters for divacancies along different crystal directions. We did not explore this latter aspect since this makes the number of fitting parameters very high and we analyzed that the results would not have statistical significance for our data set. For non-zero field we thus present ISC rates that are an average over the different defect orientations.


We thus carried out a fitting approach where we used six global ISC rates that can vary as a function of field, while maintaining single parameters $D_e$, $E_e$, $G_0$ and $\mathcal{E}_0$ for the full data set. To suppress statistical fluctuations we effectively fit average ISC rates over magnetic field ranges of 10~mT width in Fig.~\ref{Fig:FigPLEspec}a (over which the spins states do not strongly vary). We then see ISC rates that show a weak variation with field, and become constant at magnetic fields over 40~mT. This can be understood by noting that this is beyond any anticrossings in the $\ket{g_i}$ and $\ket{e_j}$ levels where the spin states mix. Here, the spin states become field independent and this results indeed in stable ISC rates from the fitting.

The results for the ISC rates are presented in Table~\ref{Tab:ISCrates} for the applied field near $B=0~{\rm mT}$ and $B>40~{\rm mT}$ (the approach with field-independent ISC rates gave values between these two cases, with a least-squares error from fitting that was 20\% higher).
The result with these ISC rates are also presented in Fig~\ref{Fig:FigPLEspec}e.
At low fields, these ISC rates indicate relatively strong values for ISC out of levels $\ket{e_m}$ and $\ket{e_u}$, and preferential ISC decay into the level $\ket{g_l}$. This identifies how single-laser excitation into the phonon sideband
or driving the ZPL with spectrally broad laser pulses (which both give parallel excitation from all spin levels, and is for example applied in ODMR studies \cite{SI-koehl2011nature,SI-falk2013natcomm}) results in spin polarization in the ground state.


\begin{table}[h!]
\centering 
\resizebox{9cm}{!}{

\begin{tabular}{|c||c|c|c||c|c|c|}
\hline     &$X^g_l $     &$X^g_m$      &$X^g_u$  &$X^e_l$     &$X^e_m$      &$X^e_u$ \\
\hline $B\approx 0 {\rm mT}$
   &14$\pm$9  &4$\pm$3  &6$\pm$3 &2$\pm$1  &6$\pm$3  &5$\pm$3  \\
   $B >40 {\rm mT}$
   &3$\pm$3  &5$\pm$3  &1.1$\pm$0.7 &3.4$\pm$0.8  &8$\pm$3  &2.9$\pm$0.5  \\
\hline
\end{tabular}}
\caption{Fitting results for the ISC rates. All values are in MHz.} \label{Tab:ISCrates}
\end{table}


\newpage

\section{Modeling of Coherent Population Trapping}
\label{Sec:SOMmodelCPT}

We investigated the agreement between our CPT observations (Fig.~\ref{Fig:FigCPTtraces}) and the standard theoretical description for this phenomenon \cite{SI-fleischhauer2005rmp} (unlike the PLE modeling of the previous section, this modeling does include coherences between the levels). This also allows for extracting the ground-state spin dephasing time $T_2^*$ for our ensemble. For making the fits for panels I-V in Fig.~\ref{Fig:FigCPTtraces}, we approximate the six-level system as an effective three-level system, where we only account for the levels that participate in a laser-driven $\Lambda$ configuration (Fig.~\ref{Fig:FigPLEspec}d). For this system we determine the steady-state solution of the density-matrix master equation, with the approach described in Ref.~\onlinecite{SI-fleischhauer2005rmp}. In order to relate it to the measured data from our two-laser spectroscopy technique, we fit using a sum of master equation solutions, where the summation is over a range of inhomogeneous broadening values for the optical transition.
(giving that the CPT dip is always in the middle of a PLE spectral line).
In addition, we take into account the decaying power along the beam path in our multipass sample (using a decay constant that was measured), by fitting with a sum of master equation solutions with exponentially decreasing Rabi frequencies.
For the measurements that show two (four) CPT dips due to small misalignment angles for the applied magnetic field, we sum the PLE from two (four) contributions. Here each contribution is calculated for one set of misalignment angles that occur within the ensemble that is addressed (representing one of the two or four orientations within ensemble $P$ or $R$, respectively).
In order to obtain good fits we need to account for a background signal in the PLE data, in particular for PLE lines with CPT that are on the wing of a neighboring PLE line. In the calculations we account for this with a linear-in-frequency background signal. We can thus calculate the CPT line shapes that appear in the PLE data, and the results are presented as the solid lines in Fig.~\ref{Fig:FigCPTtraces}. These results give a $T_2^*$ value of 42$\pm$8~ns.

\newpage

\section{Angular dependence of CPT}
\label{Sec:SOMangleCPT}

\begin{figure}[h!]
\centering
\includegraphics[width=16cm]{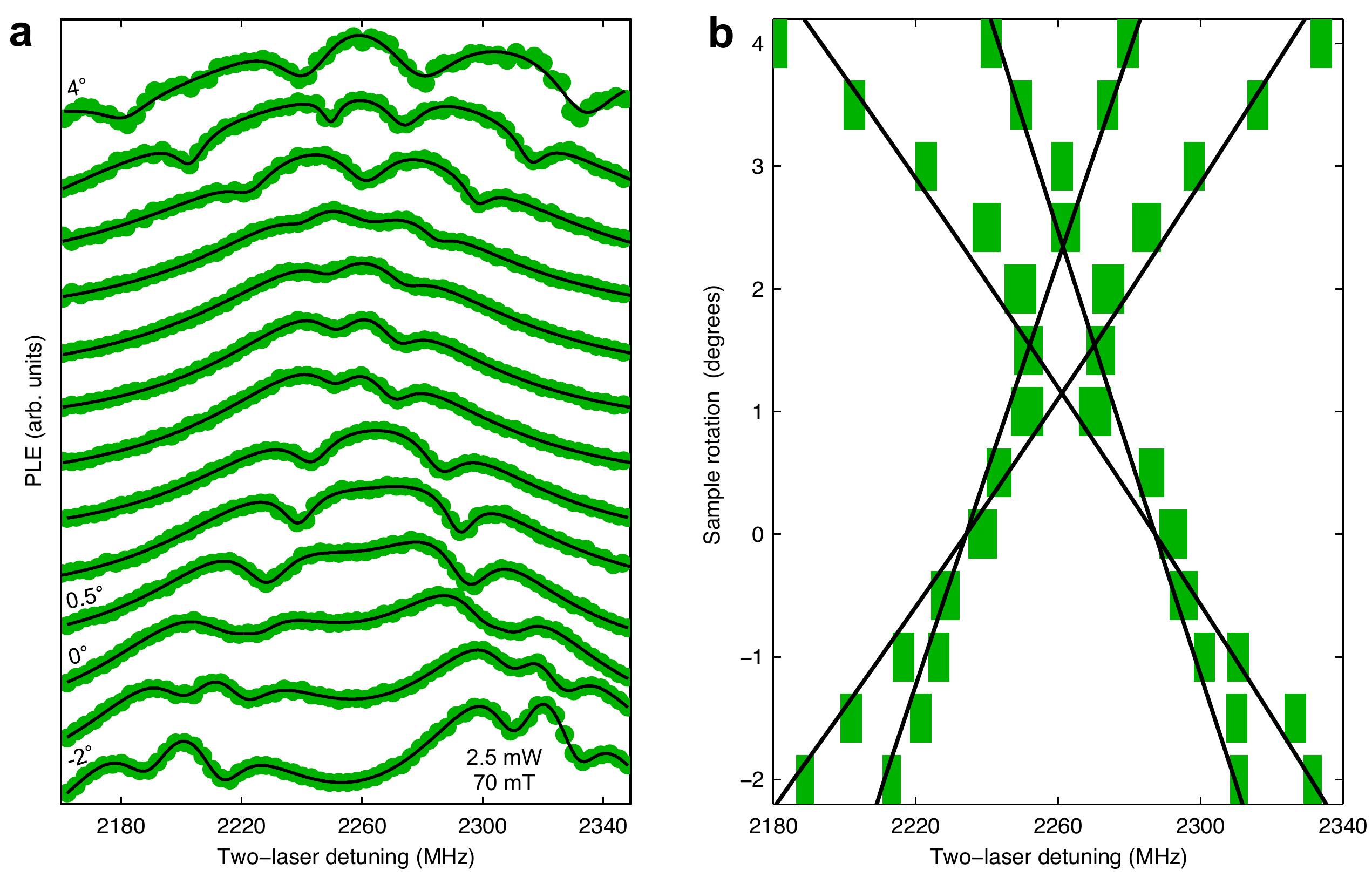}
\caption{\textbf{Sensitivity of CPT to magnetic field alignment.}
\textbf{a}, CPT in sub-ensemble $R$ at 70~mT and 2.5~mW power of the lasers, where the sample holder was rotated between measurements in steps of 0.5$^\circ$. The splitting of the CPT dip is clearly dependent on angle, increasing when the symmetry between defect orientations is reduced. Black lines are fits made with a sum of Lorentzians. \textbf{b}, CPT detunings from the fits in Fig.~\ref{Fig:Fig_CPTvsPHI}a, where the box size indicates the confidence interval. The black lines are a fit made using the ground state Hamiltonian, where the asymmetry comes from a small tilt off the sample on its holder of 1.1$^\circ$.}
\label{Fig:Fig_CPTvsPHI}
\end{figure}

To find evidence for our interpretation that the splitting of the CPT dips in panels III-V of Fig.~\ref{Fig:FigCPTtraces} is a result of minute magnetic field misalignment, we rotated the sample holder inside the cryostat in steps of 0.5$^\circ$. This is a rotation around the vertical axis of Fig.~\ref{Fig:Fig_sample}, and yields a combined variation of $\varphi$ and $\theta$ offsets due to the 45$^\circ$ forward tilt in our sample mounting, see Fig.~\ref{Fig:FigLatticePL}a and Fig.~\ref{Fig:Fig_sample}. Results from these measurement are presented in Fig.~\ref{Fig:Fig_CPTvsPHI}a for sub-ensemble $R$, with magnetic field and laser powers as indicated. The CPT dips clearly move apart as a function of sample rotation, finally showing four separate dips at large rotations. These directly reflect the four defect orientations within the $R$ sub-ensemble, with the symmetry broken by the sample rotation. Figure~\ref{Fig:Fig_CPTvsPHI}b presents the values of the two-laser detunings where the CPT dips occur (determined with phenomenological fitting of the data in Fig.~\ref{Fig:Fig_CPTvsPHI}a, the marker sizes indicate 68\% confidence intervals).
The black lines in Fig.~\ref{Fig:Fig_CPTvsPHI}b are calculated positions for the CPT dips, obtained with equation~(\ref{Eq:HamilDE}) for the four different orientations. A small misalignment in sample tilt was used as fit parameter, yielding a remaining misalignment of 1.1$^\circ$ at optimal alignment with the field (again a combination of $\varphi$ and $\theta$ offsets due to the 45$^\circ$ tilt in sample mounting).


\newpage

\section{Bleaching and charge state dynamics}
\label{Sec:SOMbleaching}

When addressing the divacancies with lasers resonant on the zero-phonon line (ZPL), the amount of PLE signal that is detected gradually decreases over time, on a timescale of minutes to hours for the laser powers we use. When keeping a single laser fixed central on the ZPL for several hours, and subsequently scanning a single laser over the entire ZPL (in a few seconds), we see a clear dip in the PLE signal (Fig.~S5).
This bleaching phenomenon for SiC divacancies is an effect also seen for NV$^-$ centers in diamond, where it has been identified as a two-photon process. In case of an NV$^-$ center, it is initially negatively charged, absorbs two photons, and subsequently an electron is lost to the conduction band\cite{SI-beha2012prl}. For the SiC divacancy the initial state is charge neutral, and two-photon absorption also causes the defect to make a transition to a different meta-stable charge state. In this charged state the wavelength for the ZPL optical transitions is shifted by over 50~nm. Applying an additional laser that excites these charged divacancies brings them back to their original charge state (again in analogy with observations on NV$^-$ centers \cite{SI-beha2012prl}). We found that applying a 300~$\mu$W repumping laser at 685~nm completely removes the bleaching dip after a few seconds, and we had such a repump laser permanently on during all the two-laser spectroscopy experiments. It is important that this laser is far-removed from the ZPL and the phonon sideband for excitation of our neutral divacancy, such that it only excites divacancies in the altered charge state. This avoids that the repump laser interferes with driving optical transitions for the neutral $V_{SiC}$ and our two-laser spectroscopy. Notably, with selectively bleaching most of the left and right wing of the inhomogeneous linewidth of the ZPL of our neutral divacancy (the inverse of the result in Fig.~S5)
one can create a long-lived sub-ensemble with a strongly reduced inhomogeneity for the optical transition.


\begin{figure}[h!]
\centering
\includegraphics[width=7cm]{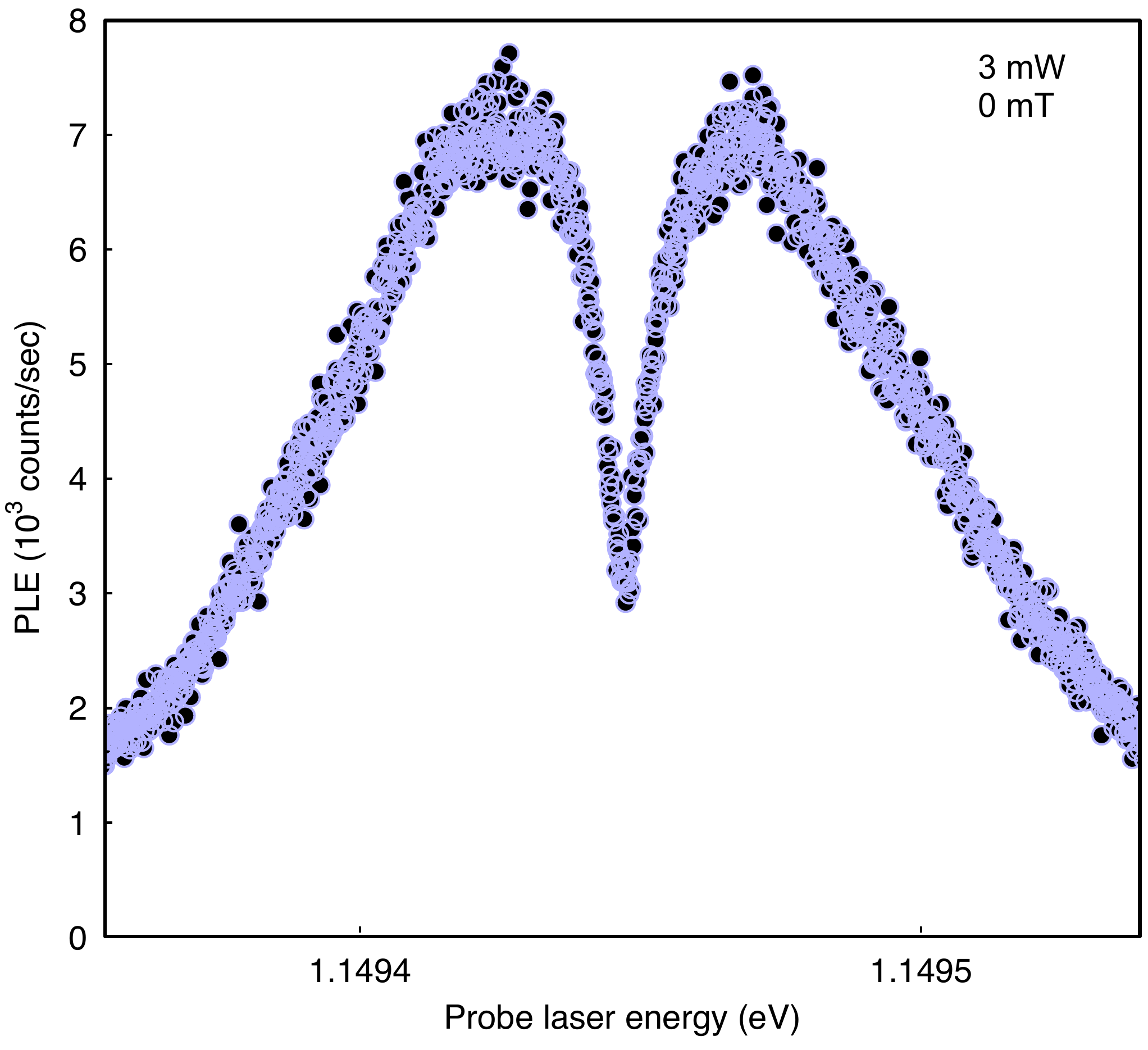}
\caption{\textbf{Charge-state bleaching.}
The inhomogeneously broadened ZPL as in the inset of Fig.~\protect{\ref{Fig:FigLatticePL}}c, with a dip of 3.5~GHz full-width (FWHM). The dip was formed after keeping a CW laser fixed at the dip frequency, causing the PLE to steadily decrease to this value in a few hours. This effect, known as bleaching, is caused by divacancies changing their charge state.
In the altered charge state the divacancies are off-resonant with the laser, and remain in this state for hours after the laser driving. This spectrum with a dip can be restored to the original peak shape (inset of Fig.~\protect{\ref{Fig:FigLatticePL}}c) within seconds by applying a 685-nm laser. This laser is exciting the charged divacancies in their phonon sideband, and this pumps them back to their neutral charge state.}
\label{Fig:FigBleaching}
\end{figure}
